# Study of gamma ray response of R404A superheated droplet detector using a *two-state* model

**P.K. Mondal** [a,†,*] **and B.K. Chatterjee** [a]

[a] Department of Physics, Bose Institute, 93/1 A P C Road, Kolkata 700009, India.
  e-mail: prasanna_ind_82@yahoo.com (P.K. Mondal)
       barun_k_chatterjee@yahoo.com  (B.K. Chatterjee)

**Abstract**
The superheated droplet detector (SDD) is known to be gamma insensitive below a threshold temperature which made them excellent candidates for neutron detection in the presence of gamma rays. Above the threshold temperature, the gamma ray detection efficiency increases with increase in temperature. In this work the gamma ray threshold temperature has been studied for SDD using R404A as the active liquid and is compared to the theoretical prediction. The temperature variation of gamma ray detection efficiency and interstate transition kinetics has also been studied using a two-state model. The experiments are performed at the ambient pressure of 1 atmosphere and in the temperature range of 17-32°C using a 662 keV $^{137}$Cs gamma ray source.

## 1. Introduction

It is well known that a liquid can be superheated to a temperature above its boiling point. The superheated state is a metastable state, where a minor perturbation like mechanical vibration, thermal fluctuation, energy deposition by energetic radiation etc., can trigger the formation of a stable vapor phase. Application of superheated liquid in radiation detection dates back to the time of the bubble chamber invented by Glaser [1]. The use of a superheated liquid in a more convenient form of radiation detector, known as the superheated drop detector, was reported by Apfel [2]. In superheated droplet detector (SDD), the whole liquid is dispersed in the form of minute droplets in a gel medium. SDDs are used as radiation detector in various areas, from health physics [3] to high energy physics [4]. Superheated liquid is also considered as a suitable detector to search for dark matter [5-7], since by choosing the operating temperature and pressure it can be made sensitive to high ionizing radiations yet virtually insensitive to the majority of backgrounds like gamma rays, beta rays, cosmic muons etc. PICASSO [5], SIMPLE [4,6] and COUPP [7] are the groups working on the dark matter detection using superheated liquids. For the preparation of SDD different low boiling point liquids are used, such that at their operating temperature they are in reasonably superheated state and can be used for detection of ionizing radiations. Recent studies also indicated that in superheated emulsion there exist two groups of droplets one having much shorter lifetime than the other [8,9]. The decay of these droplets in SDD has been modeled earlier [9] using a *two-state* model which explains the ambiguity in the nucleation rate data. In this paper we have studied the characteristics of R404A (nearly azeotropic blend of $C_2H_3F_3$, $C_2H_2F_4$ and $C_2HF_5$; b.p. -46.5$^o$C) based SDD, which is gamma ray sensitive at room temperature. The gamma ray detection threshold temperature of R404A is obtained using a 662 keV $^{137}$Cs gamma ray source. The nucleation parameters, detection efficiency and interstate transition kinetics of the two metastable state is studied using the *two-state* model [9]. To the best of our knowledge R404A based SDD has not been studied earlier.

---

[†] Present address: Astroparticle Physics and Cosmology Division,
    Saha Institute of Nuclear Physics, 1/AF Bidhannagar, Kolkata 700064, India.
       e-mail: prasanna.mondal@saha.ac.in.

## 2. Theory

In the superheated state of a liquid, the liquid is under tension and at a temperature higher than its boiling point at a given pressure. Here the liquid teams with a dynamic population of microbubbles [10], which pool the energy from the liquid into pockets where the tear grows (spherically) to a maximum size before imploding back and vanishing. A barrier [10] (in the radial coordinate) due to the interplay between the surface and volume energy governs the stability of the microbubbles. Occasionally a microbubble will have sufficient energy to overcome the barrier to cause spontaneous homogeneous nucleation. Nucleation may also be triggered by energetic radiation which deposits energy in the liquid and causes radiation induced nucleation. The frequency of spontaneous homogeneous nucleation is quite low compared to the induced nucleation, and this enables one to use SDD as a radiation detector.

During the bubble nucleation, after overcoming the barrier, the microbubble grows by the evaporation of the superheated liquid. This growth of microbubble and subsequent droplet vaporization is accompanied by the emission of an acoustic pulse [11]. Each individual nucleation is accompanied by the emission of an acoustic pulse and a change in volume. The acoustic pulse or the change in volume can be detected electronically [11-13], which enables one to detect the nucleation.

Superheated droplets, when exposed to energetic radiation, are expected to decay monotonically. However, when the droplets are irradiated multiple times with a radiation-off period in between successive irradiations, it is observed that the nucleation rate at the beginning of the irradiation considerably increases compared to the rate at the end of the previous irradiation [8,9]. This discrepancy in the nucleation rate data indicates that in SDD the droplets are in two metastable states. The droplets continuously move from one of these states to the other and these two states are in thermal equilibrium with each other. When the droplet is in the *normal metastable state* (**N**), it has much longer lifetime than when it is in the *second metastable state* (**S**). These two states were observed for various superheated liquids when irradiated with neutrons and gamma rays [8,14,15].

At thermal equilibrium the droplets in the detector are distributed among the two metastable states. When irradiated the short lived droplets in *second metastable state* decay much faster than the others in *normal metastable state*, giving a sharp initial fall in the nucleation rate, after which only the long-lived droplets remain and decay [8,9]. The short lived droplets repopulate from the *normal metastable state* during the radiation-off period resulting in an increase in the nucleation rate at later irradiations. It was observed that $d$, the transition rate from second metastable state to normal metastable state ($S \rightarrow N$), is larger than the transition rate from normal metastable state to second metastable state ($N \rightarrow S$) $c$ [9].

During irradiation the superheated droplets decay due to induced and spontaneous nucleations and the nucleation frequency of the droplets in normal and second metastable states can respectively be expressed as [9],

$$b = b^{spont} + b^{induced} = k_o v + k_1 v \psi \quad (1)$$

and

$$a = a^{spont} + a^{induced} = q_o v + q_1 v \psi. \quad (2)$$

Here $v$ is the droplet volume, $\psi$ is the radiation flux, $k_o$, $q_o$ are the spontaneous nucleation rate per unit volume for the normal and second metastable states respectively and $k_1$, $q_1$ are the radiation induced nucleation frequency per unit volume per unit flux for the normal and second metastable states respectively. The first term on the right hand side of Eqs. (1) and (2) accounts for the spontaneous nucleation and the second term accounts for the radiation induced nucleation.

The nucleation parameters of the two metastable states $k_o$, $q_o$, $k_1$ and $q_1$, and the interstate transition rates $c$ and $d$ can be obtained by fitting the multi-exposure nucleation rate data [9]. In this work it is done by using the *two-state* model and the temperature variations of the parameters $k_o$, $q_o$, $k_1$, $q_1$, $c$ and $d$ are obtained for R404A detector.

**2.1. Detection efficiency**

It is well known that for gamma rays the energy deposition takes place by the secondary electrons, having much larger ranges compared to that of the ions. The detection efficiency of the detector ($\eta_{det}$) is defined as the ratio of the number of nucleation events recorded to the number of energetic particles incident on the detector and can be expressed as,

$$\eta_{det} = \frac{\text{no. of counts received}}{\text{no. of particles incident on the detector}}. \quad (3)$$

In case of SDD the detection efficiency can be expressed as [15,16],

$$\eta_{det} = \frac{-(dS/dt)_{t=0}}{\psi A}. \quad (4)$$

Here, $dS/dt$ is the rate of nucleation of the superheated droplets, $\psi$ is the radiation flux and $A$ is the sagittal sectional area of the detector. Considering the presence of two metastable states Eq. (4) can also be written as,

$$\eta_{det} = \frac{-(dS/dt)^N_{t=0}}{\psi A} + \frac{-(dS/dt)^S_{t=0}}{\psi A}. \quad (5)$$

Here the first and second terms on the right hand side of Eq. (5) correspond to normal and second metastable state respectively. Since the spontaneous nucleation rate is negligibly small, Eq. (5) can be expressed as,

$$\eta_{det} = \frac{k_1 S_o^N \bar{v}}{A} + \frac{q_1 S_o^S \bar{v}}{A} \qquad (6)$$

where, $\bar{v}$ is the average volume of the droplet, $S_o^N = S_o c/(c+d)$ is the number of normal metastable drops initially present in the detector and $S_o^S = S_o d/(c+d)$ is the number of second metastable drops initially present in the detector, $S_o$ being the total number of drops initially present in the detector. Hence the detection efficiency of the detector is

$$\eta_{det} = \frac{(k_1 c + q_1 d) S_o \bar{v}}{A(c+d)} = \frac{(k_1 c + q_1 d) V}{A(c+d)} \qquad (7)$$

here $S_o \bar{v} = V$ is the total volume of the active liquid initially present in the detector. Thus the detection efficiency per unit volume of the active liquid is expressed as

$$\eta_D = \frac{(k_1 c + q_1 d)}{A(c+d)} \qquad (8)$$

Eq. (8) is used for the measurement of gamma detection efficiency of the detector.

### 3. Detector fabrication

In order to get stable superheated droplet detector, one has to minimize the heterogeneous nucleation in superheated emulsion. Some essential criteria for preparing a stable SDD are: (i) ideally the gel medium should be free from all heterogeneous nucleation sites, (ii) the active liquid must be immiscible in the gel, (iii) the holding medium should have the proper viscosity and shear elasticity in order to keep the droplets suspended in the holding medium, while preventing them from settling down, and yet not so high as to prevent the vapor bubbles from escaping from the gel matrix.

The viscoelastic gel used in SDD is prepared by mixing glycerol with commercial ultrasound gel in suitable proportion such that it can hold the droplets in suspension. The holding medium provides smooth surface and reduces the possibility of heterogeneous nucleation due to the container surface roughness. As stated earlier, dissolved gas and air pockets in the gel act as heterogeneous nucleation sites and reduce the stability of the metastable liquid. To remove them the gel is thoroughly degassed.

Another important criterion which should be taken into account at the time of preparation of gel is the solubility of refrigerant liquid in gel. The more soluble the liquid is, the greater is the diffusion of the liquid into the gel, enhancing Ostwald ripening which gradually vanishes the droplets below a critical radius, while, enlarging those above it [17,18]. This reduces the number of liquid droplets in SDD. It was also observed that the R404A droplets are unstable in the emulsions due to their high solubility in water (0.19 wt% at 1 bar and 25$^o$C) and the addition of surfactant produces much stable samples by reducing the Ostwald ripening rate [19-21]. In this work we have used Tween 80 as the surfactant, which was added to the gel (0.1 % of the gel by volume) before the emulsification.

In the emulsification process initially about 50 ml of freon R404A is transferred to about 100 ml of the degassed gel in a pressure tight container and then sheared with the help of an electric stirrer, which breaks the liquid into small droplets. After shearing the droplets are brought to the superheated state by slowly releasing the container pressure. The emulsion is then poured into 15 ml glass vials, each having about 10 ml of emulsion. Since R404A has a very low boiling point, during liquid transfer and pressure release process most of the active liquid vaporizes and at the end of the fabrication the emulsion usually contains very small amount of active liquid which is less than 5% of the total volume of the emulsion. The emulsion obtained by this method is polydisperse in nature. The droplet size distribution of the detector was measured using a new technique [22] and is shown in Fig. 1. It is observed that in R404A SDD the droplet size varies in the range of 25-125 µm. In Fig. 1 $f(v)$ represents the normalized droplet volume distribution of the emulsion, which is used in fitting the nucleation rate data [9].

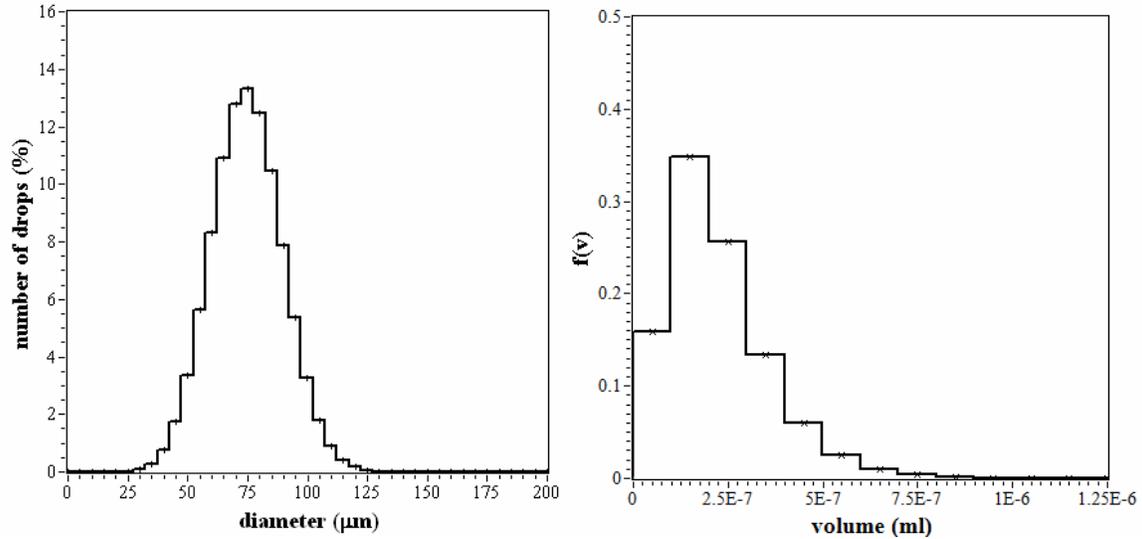

**Fig. 1** Droplet size distribution in R404A detector.

### 4. Experimental methods

For the characterization of R404A detectors two set of experiments are performed. First we have studied the temperature variation of the response of R404A detector to the gamma rays by using a $^{137}$Cs gamma ray source. Then we have performed other experiments for studying the nucleation parameters of the R404A detector using the $^{137}$Cs source.

### 4.1. Experiment for study the gamma ray sensitivity of the R404A SDD

The vaporization of a superheated droplet is associated with the generation of an acoustic pulse, which can be detected by a piezoelectric transducer that converts it into electric signal. The electric pulses can be converted into TTLs with the help of a pulse shaping circuitry. The present investigation is based on an active device [23] which is sensitive to individual bubble nucleation in superheated droplet. From this active device one obtains the nucleation rate data, i.e. the number of drops ($N(t)$) vaporized during the dwell-time at time $t$.

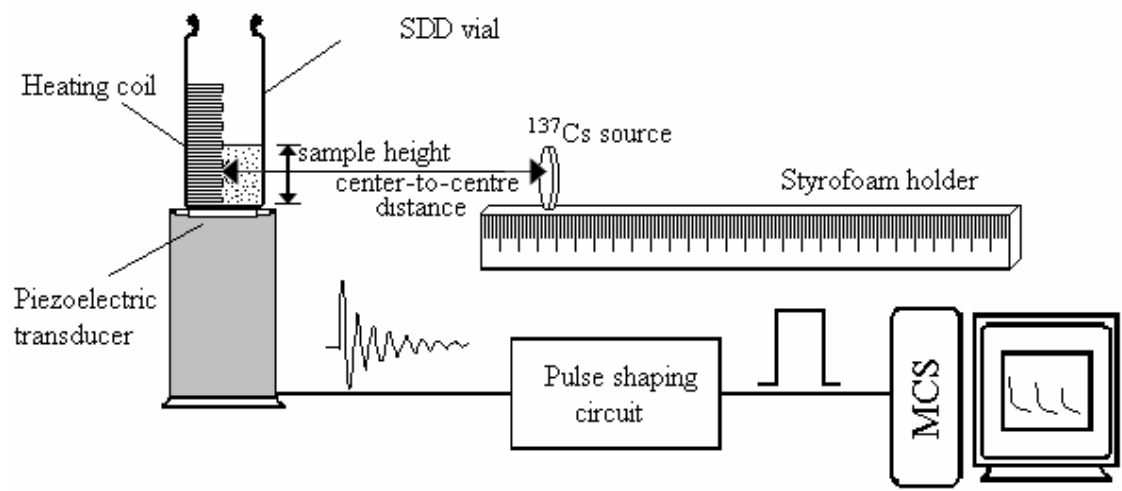

**Fig. 2** Schematic diagram of the experimental setup used for R404A experiments using a $^{137}$Cs gamma ray source.

To obtain the gamma ray detection threshold temperature, the R404A detector was exposed to the gamma rays at a gamma ray flux of $2.1\times10^5$ cm$^{-2}$s$^{-1}$, while the temperature of the detector was increased continuously in small steps ($\Delta T \approx 0.2^{\circ}$C/min). A schematic of the experimental setup is shown in Fig. 2. In this experiment the detector vial was placed on the top of the piezoelectric transducer coupled by an ultrasonic gel. The detector vial was wrapped with a heating coil which was connected to a variac, by which the temperature of the detector was controlled. The droplet nucleation was detected by the active device and the TTL pulse counts were collected as a function of time using a MCS. In this method the number of drops vaporized in a dwell time of 30 s was acquired as a function of temperature. A similar experiment was also performed in which the nucleation rate data was acquired in absence of the gamma ray source to obtain the temperature variation of the spontaneous nucleation events (which also includes the events for the background radiations). The temperature variations of the normalized response of R404A obtained with $^{137}$Cs source and without the source are shown in Fig. 3. Since the number of drops decreases with time, the nucleation rate is normalized with respect to the surviving number of droplets present at any given time. To obtain the response separately for gamma ray induced events, the spontaneous nucleation data was subtracted from the data obtained with $^{137}$Cs source (Fig. 4). From Fig. 4, it is observed that the R404A liquid becomes sensitive to 662 keV gamma rays at about 13.3$^{\circ}$C. The obtained gamma ray threshold for $^{137}$Cs is in good agreement with the proposed empirical model by d'Errico et al. [24,25], which predicts the threshold to be immediately above the mid-point between boiling and critical temperatures of the halocarbons. According to this model the threshold for R404A, having the boiling point and critical point of -46.5$^{\circ}$C and 72.1$^{\circ}$C respectively, would be at about 14$^{\circ}$C.

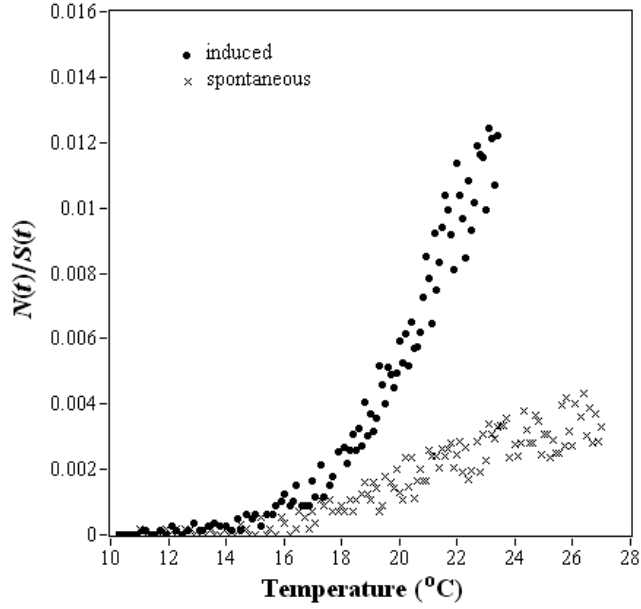

**Fig. 3** Temperature variation of the normalized nucleation rate of R404A SDD obtained with $^{137}$Cs gamma ray source and without the gamma ray source.

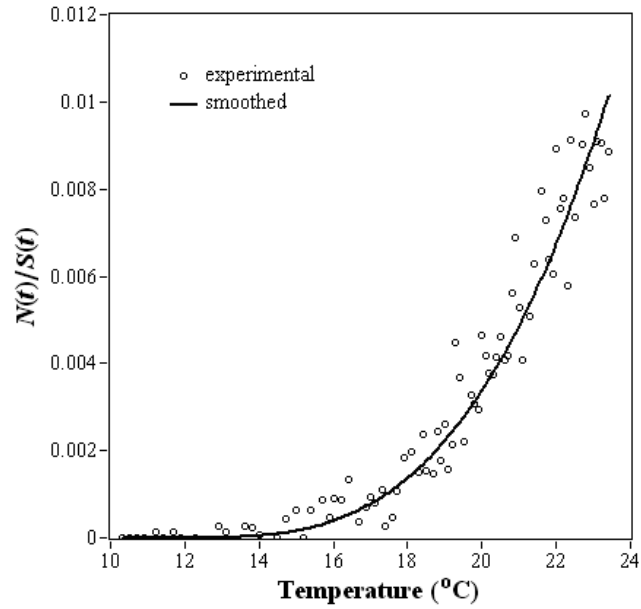

**Fig. 4** Temperature variation of the response of R404A SDD for only 662 keV gamma rays from $^{137}$Cs gamma ray source.

**4.2 Experimental setup for the study of two metastable states**

To measure the parameters of the two metastable states multi-exposure nucleation rate data is required [9]. The experimental setup is similar to that used for the determination of gamma ray threshold temperature of the detector. Here at a constant temperature and at ambient pressure the detector was periodically irradiated with gamma rays with some periods of radiation-off regions in between two successive irradiations. In these experiments initially the detector was irradiated with gamma rays for a few minutes. The radiation was turned off for a period of time (by removing the source) and then turned back on again by placing the source. Such switching off and on the irradiation was repeated again where the radiation-off periods were varied.

To understand rates of interstate transition and also for obtaining the temperature variations of spontaneous and induced nucleation rates of the two metastable states experiments were performed in the temperature range of 17-32°C. In all these experiments the R404A SDD was irradiated 3 times with 662akeV gamma rays from $^{137}$Cs gamma ray source. A typical multi-exposure experimental data at temperature 29°C is shown in Fig. 5.

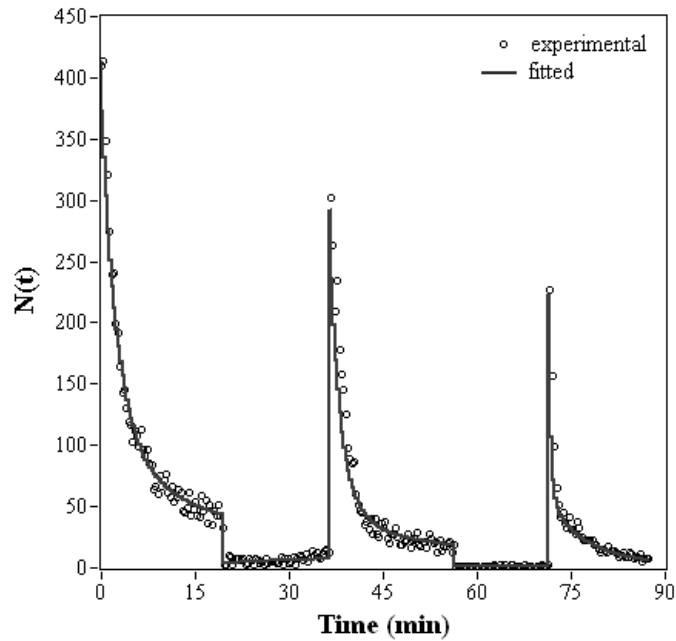

**Fig. 5** A typical experimental and fitted multi-exposure nucleation rate data for R404A SDD at 29°C.

### 5. Results and discussion

The multi-exposure nucleation rate data are analyzed using the two-state model [9]. Since the lifetime of the droplet ($b^{-1}$, $a^{-1}$ (Eqs. 1-2) is dependent on the droplet volume $v$, the volume distribution $f(v)$ plays an important role in the fitting of the nucleation rate data [23,26]. The details of the data fitting method has been reported earlier [9], where the volume distribution $f(v)$ was incorporated in the model in such a way that its evolution, due to the continuous droplet depletion, can also be included throughout all the regions of the experimental data. In this work, we have used the measured droplet volume distribution $f(v)$ which is shown in Fig. 1. By fitting a nucleation rate data one obtains the parameters $k_o$, $q_o$, $k_1$, $q_1$, $c$ and $d$. From these parameters one can obtain the other quantities like the equilibration time ($1/(c+d)$), detection efficiency ($\eta_D$) and the population of the droplets in different metastable states. A typical experimental and fitted data for R404A SDD is shown in Fig. 5. For all the multi-exposure data acquired at different temperatures these parameters are obtained and the temperature variations of the parameters are shown in Figs. 6-11. From these results it is observed that for R404A SDD detector the temperature variation of a given parameter have some characteristic features.

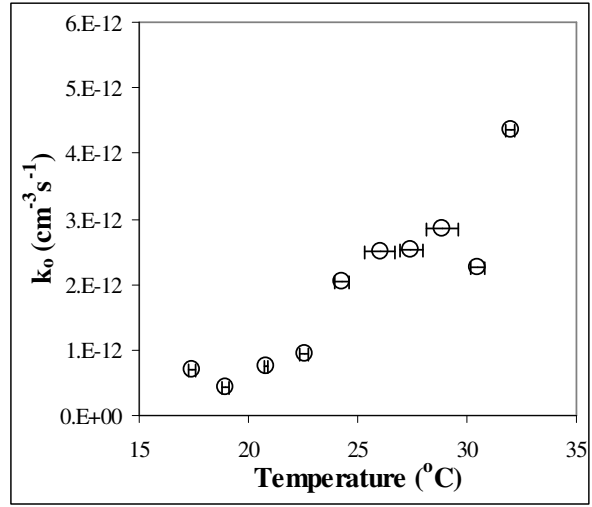

**Fig. 6** Temperature variation of $k_o$, the spontaneous nucleation frequency per unit volume for normal metastable state.

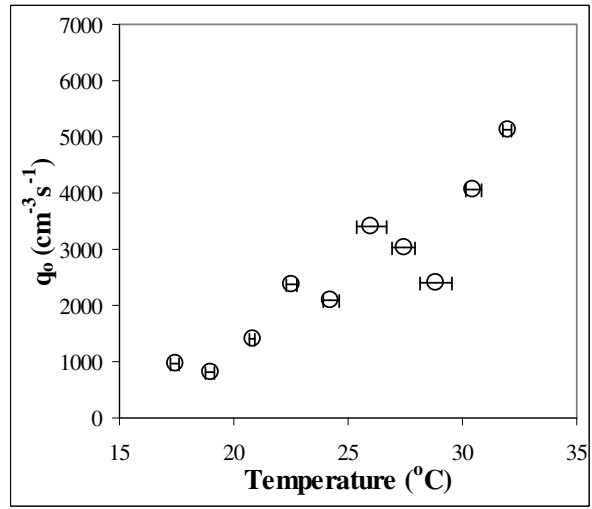

**Fig. 7** Temperature variation of $q_o$, the spontaneous nucleation frequency per unit volume for second metastable state.

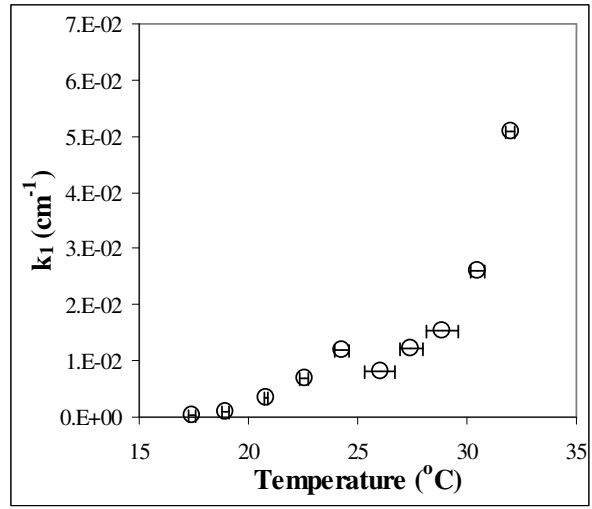

**Fig. 8** Temperature variation of $k_1$, the gamma ray induced nucleation frequency per unit volume per unit flux for normal metastable state.

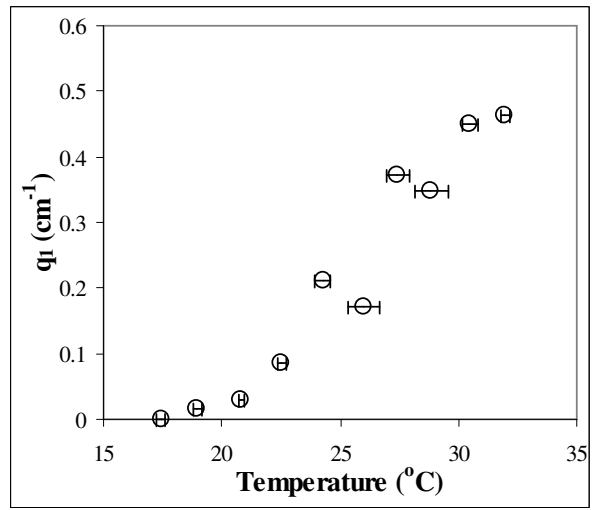

**Fig. 9** Temperature variation of $q_1$, the gamma ray induced nucleation frequency per unit volume per unit flux for second metastable state.

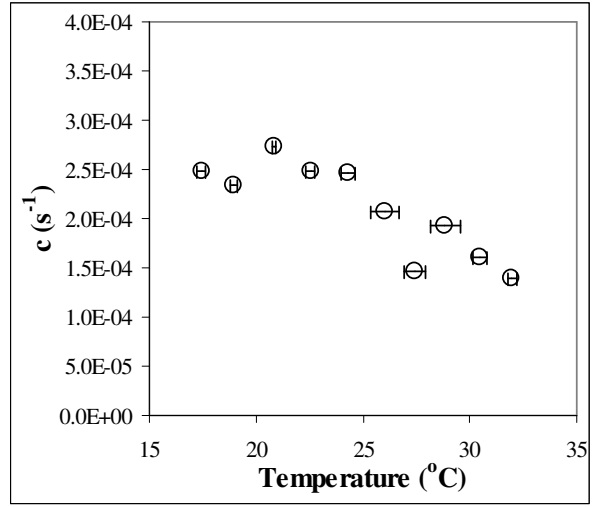

**Fig. 10** Temperature variation of $c$, the transition rate from normal metastable state to second metastable state.

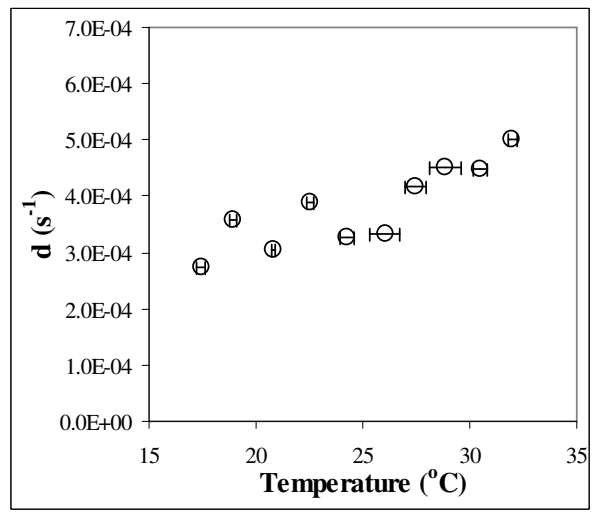

**Fig. 11** Temperature variation of $d$, the transition rate from second metastable state to normal metastable state.

The temperature variations of $k_o$ and $q_o$ are shown in Fig. 6 and Fig. 7 respectively. Here $k_o$ and $q_o$, the spontaneous nucleation frequency per unit volume of the active liquid for normal and second metastable states respectively, increase with increase in temperature. This happens due to the fact that with increase in temperature the threshold energy for nucleation ($W$) (Fig. 12) decreases and the number of microbubbles per unit volume of the liquid increases. For these two reasons with increase in temperature the probability of a microbubble to cross the nucleation barrier increases, resulting in an increase in $k_o$ and $q_o$. In Fig. 12, the threshold energy for nucleation ($W$) [10] has been obtained using the relation $W = [16\pi\gamma^3]/[3(P_v - P_o)^2]$, where $\gamma$ is surface tension of the liquid, $P_v$ is the equilibrium vapour pressure and $P_o$ is the ambient pressure.

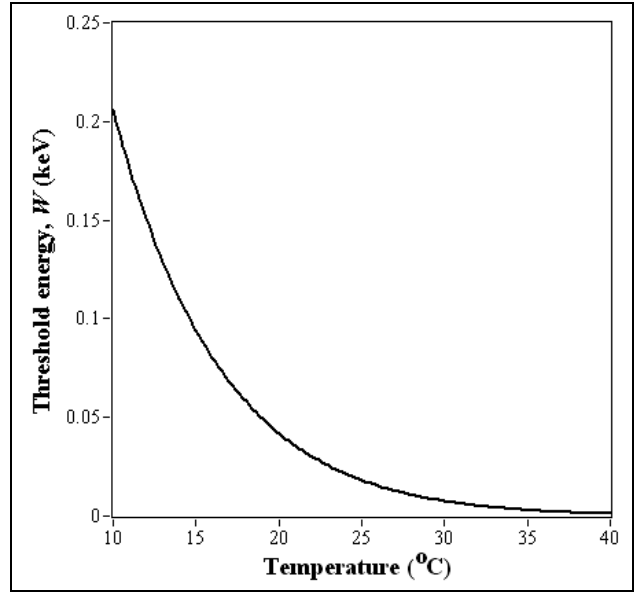

**Fig. 12** Temperature variation of the threshold energy for nucleation ($W$) for R404A.

The radiation induced nucleation frequencies of the two metastable states also increase with increase in temperature, as shown in Figs. 8-9. Here $k_1$ and $q_1$ are the radiation induced nucleation frequencies per unit flux per unit volume of the active liquid for normal and second metastable states respectively. For radiation induced nucleation, in addition to the threshold energy and microbubble density, the interaction cross-section and the energy deposition affect the nucleation rate. As the threshold energy for nucleation decreases with rise in temperature, more and more electrons contribute to the droplet nucleation. Hence the induced nucleation frequencies $k_1$ and $q_1$ increase with temperature. Here it is observed that the spontaneous nucleation frequencies for the second kind of drops are much higher than that for the normal drops.

The temperature variations of the transition rate $c$ and $d$ are shown in Fig. 10 and Fig. 11 respectively. It is observed that with rise in temperature, $c$ decreases, while $d$ increases. It implies that, as the temperature increases the probability of $\mathbf{N} \rightarrow \mathbf{S}$ transition decreases while the probability of $\mathbf{S} \rightarrow \mathbf{N}$ increases. For this reason the population of droplets in second metastable state decreases with increase in temperature (Fig. 13). At equilibrium the fraction of droplets present in the second metastable state is,

$$\mathbf{P_S} = \frac{S_o^S}{S_o} = \frac{c}{c+d}. \qquad (9)$$

Using $c$ and $d$, the temperature dependence of $\mathbf{P}_s$ is obtained and is shown in Fig. 13. Here it is observed that as the temperature increases the droplet population in the short lived state decreases.

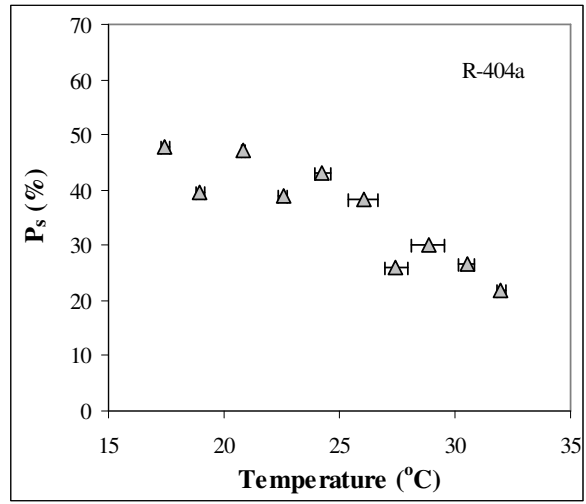

**Fig. 13** Temperature variation of the short-lived droplet population in R404A detector.

Using $c$ and $d$ the equilibration time of the system can be estimated. In absence of decay after a time $\tau = 1/(c+d)$, known as the equilibration time, the system will reach equilibrium, and after this time there will be no substantial change in the population of the droplets present in different metastable states. The temperature variation of $\tau$ is shown in Fig. 14. Here it is observed that as the temperature increases the droplet populations in normal and second metastable states equilibrate among themselves in shorter times though the population of short-lived droplets decreases with increase in temperature as evident from Fig. 13.

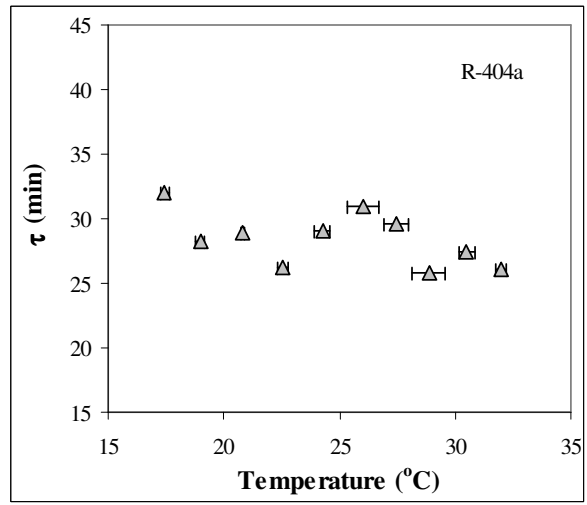

**Fig. 14** Temperature variation of the equilibration time.

The gamma ray detection efficiency ($\eta_D$) of the detector is obtained using Eq. (8). The temperature variation of $\eta_D$ is shown in Fig. 15, which shows that $\eta_D$ increases with increase in temperature.

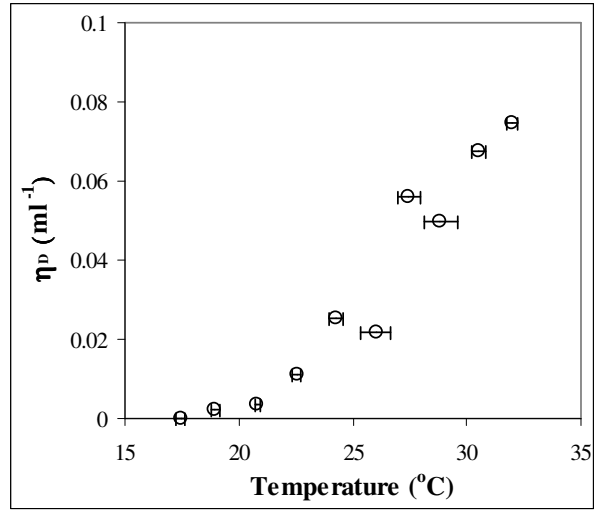

**Fig. 15** Temperature variation of gamma ray detection efficiency $\eta_D$ of R404A detector.

## 5. Conclusion

In this work we have observed that R404A is gamma ray sensitive at and above 13.3°C at one atmospheric pressure. This agrees with the empirical model by d'Errico *et al.* [24]. We also observed that R404A shows a strong presence of the second metastable state. The equilibration time is large and the proportion of the droplets in the second metastable state for R404A is also quite large in the range 17-32°C. The large value of $k_1$ and $q_1$ gives rise to a large value of the detection efficiency per unit volume of the active liquid for R404A SDD and thus makes it an ideal candidate to detect gamma rays.